%% file: lb_dis2001.tex
\documentclass{ws-p8-50x6-00}

\input{definitions}

\begin{document}

\setlength{\unitlength}{1.0mm}
\lefthyphenmin=2
\righthyphenmin=3

\title{Implications of the Tevatron Jet Results on PDF}

\author{Levan Babukhadia \\ for the \dzero\ Collaboration}

\address{Department of Physics, State University of New York\\
         Stony Brook, NY 11794, USA\\
	 E-mail: blevan@fnal.gov}


\maketitle

\abstracts{
  We report a new measurement of the pseudorapidity (\peta) and transverse-energy 
  (\et) dependence of the inclusive jet production cross section in \ppbar\ 
  collisions at $\sqrt{s}=1.8$~TeV using 95 \ipb\ of data collected with
  the \dzero\ detector at the Fermilab Tevatron. The differential 
  cross section \mycs\ is presented up to $\aeta=3$, significantly extending 
  previous measurements. 
  The results are in good overall agreement with next-to-leading order 
  predictions from QCD, indicate a preference for certain parton 
  distribution functions, and provide the world's best constraint
  on the gluon distribution at high parton momentum fraction $x$.
}

\section{Introduction}
\noindent
This past decade has witnessed impressive progress in both the theoretical 
and experimental understanding of the collimated streams of particles or ``jets'' 
that emerge from inelastic hadron collisions.
Theoretically, jet production in hadron collisions is understood within 
the framework of Quantum Chromodynamics (QCD), as a hard scattering of the 
constituent partons (quarks and gluons) that, having undergone a collision, 
manifest themselves as jets in the final state.
QCD predicts the amplitudes for the hard scattering of partons at high momentum 
transfers.
Perturbative QCD calculations of jet cross sections,~\cite{theory} using accurately 
determined parton distribution functions (PDFs),~\cite{PDFs} have 
increased the interest in jet measurements at the $\sqrt{s}=1.8$ TeV Tevatron 
proton-antiproton collider.
Consequently, the two Tevatron experiments, \dzero\ and CDF,
have served as prominent arenas for studying hadronic jets.

\begin{figure}[!ht] \centering
  \begin{picture}(118,59)  

  \put(13,20){\makebox(0,0)[rb]{\scalebox{0.9}{ \bf (a) }}}
  \put(74,20){\makebox(0,0)[rb]{\scalebox{0.9}{ \bf (b) }}}

  \put(-15,-35){\begin{picture}(59,59)
	      \epsfxsize=8.3cm 
	      \epsfysize=13.0cm 
	      \epsfbox{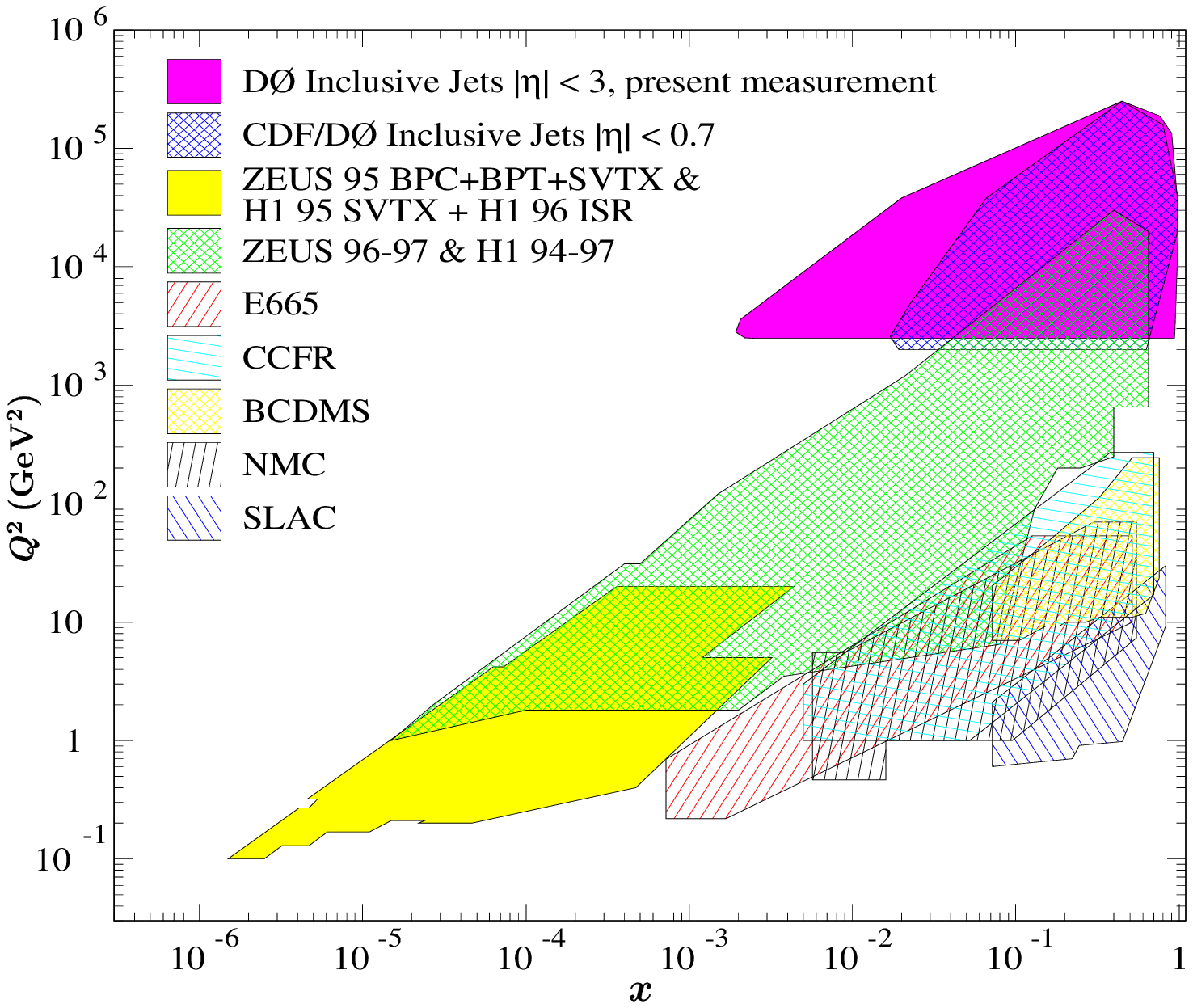}
	    \end{picture}}

  \put(44,-29){\begin{picture}(59,59)
	      \epsfxsize=8.5cm 
	      \epsfysize=11.7cm 
	      \epsfbox{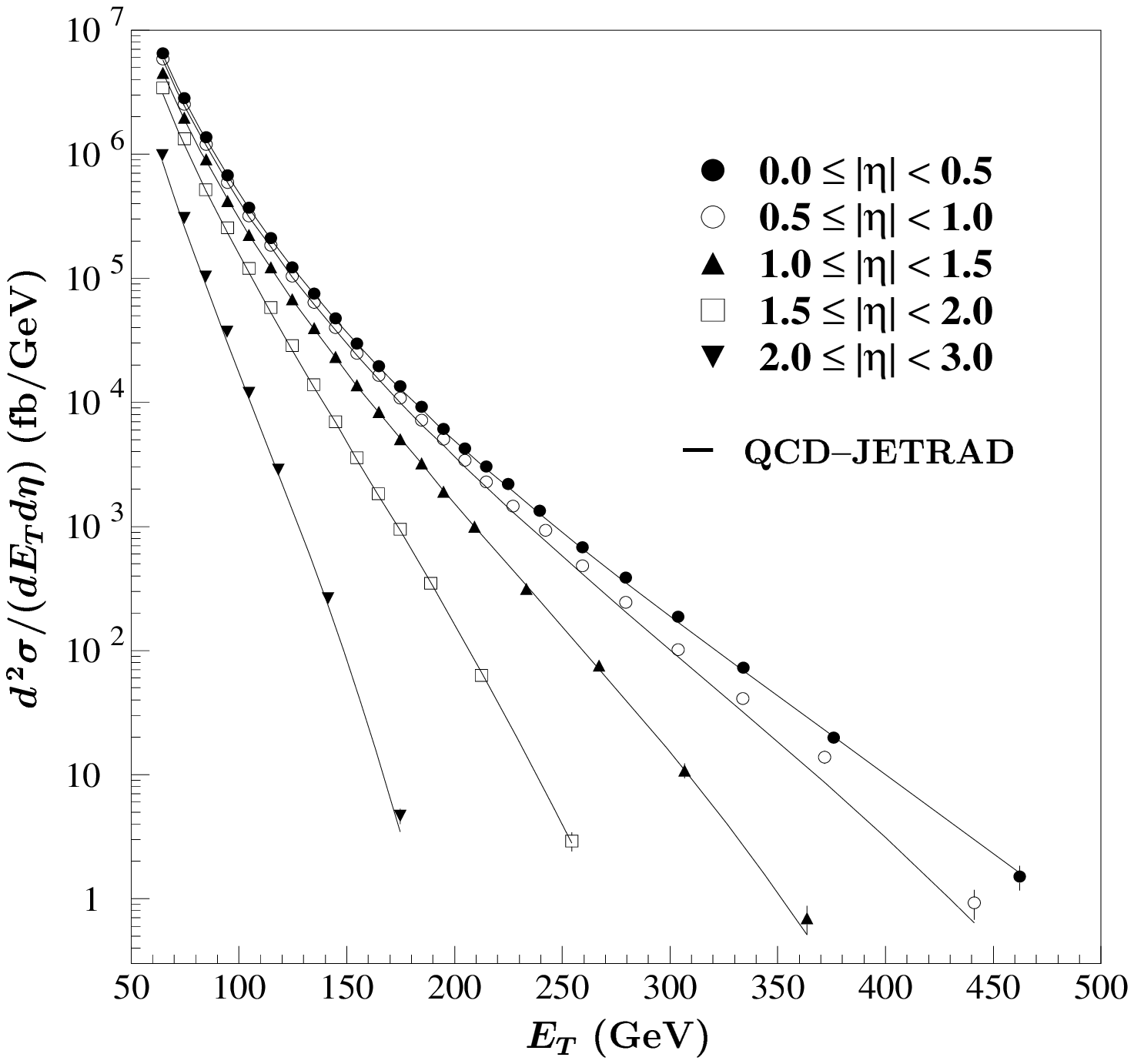}
	    \end{picture}}

  \end{picture}
  \vspace{-0.7cm}
  \caption{{\bf (a)} The kinematic reach of this measurement along with that of 
	   other collider and fixed-target experiments in the plane of 
	   the parton momentum fraction $x$ and the square of the 
	   momentum transfer $Q^{2}$, and {\bf (b)} the single inclusive jet 
	   production cross section as a function of jet \et, in five 
	   pseudorapidity intervals, showing only statistical uncertainties, 
	   along with theoretical predictions.}	 
  \label{fig:1}
  \vspace{-0.7cm}
\end{figure}

Here, we report a new measurement of the pseudorapidity (\peta) and transverse-energy 
(\et) dependence of the inclusive jet production cross section,~\cite{myprl}
which examines the short-range behavior of QCD, the structure of the proton in 
terms of PDFs, and possible substructure of quarks and gluons.
We present the differential cross section \mycs\ as a function of jet \et\ in 
five intervals of \peta, up to $\aeta=3$, where the pseudorapidity is defined 
as $\eta=\ln \left[ \cot \left( \theta/2 \right) \right]$, with $\theta$ 
being the polar angle.
The present measurement is based on 95~\ipb\ of data collected with the 
\dzero\ detector~\cite{D0detector} during 1994--1995, and significantly 
extends previous measurements,~\cite{PDG} as indicated by the
kinematic reach shown in Fig.~\ref{fig:1}a.

\section{Jet Definition and Experimental Systematics}
\noindent
Jets are reconstructed using an iterative 
cone algorithm with a fixed cone radius of $\mathcal{R}=0.7$ in 
\etaphi\ space, where $\varphi$ is the azimuth.
Offline data selections eliminate contamination from background 
caused by electrons, photons, noise, or cosmic rays.
This is achieved by applying an acceptance cutoff on the 
\z--coordinate of the interaction vertex, flagging events with large 
missing transverse energy, and applying jet quality criteria.
Details of data selection and corrections due to noise and/or 
contamination are described elsewhere.~\cite{myprl}
A correction for jet energy scale accounts for instrumental effects 
associated with calorimeter response, showering and noise, as well as 
for contributions from spectator partons, and corrects on average the
reconstructed jet \et\ to the ``true'' \et.~\cite{myprl,d0jes}
The effect of calorimeter resolution on jet cross section is removed 
through an unfolding procedure.~\cite{myprl}

\begin{figure}[!ht] \centering
  \begin{picture}(118,85)  

  \put(-15.5,-27){\begin{picture}(59,85)
	      \epsfysize=14.0cm 
	      \epsfxsize=8.6cm 
	      \epsfbox{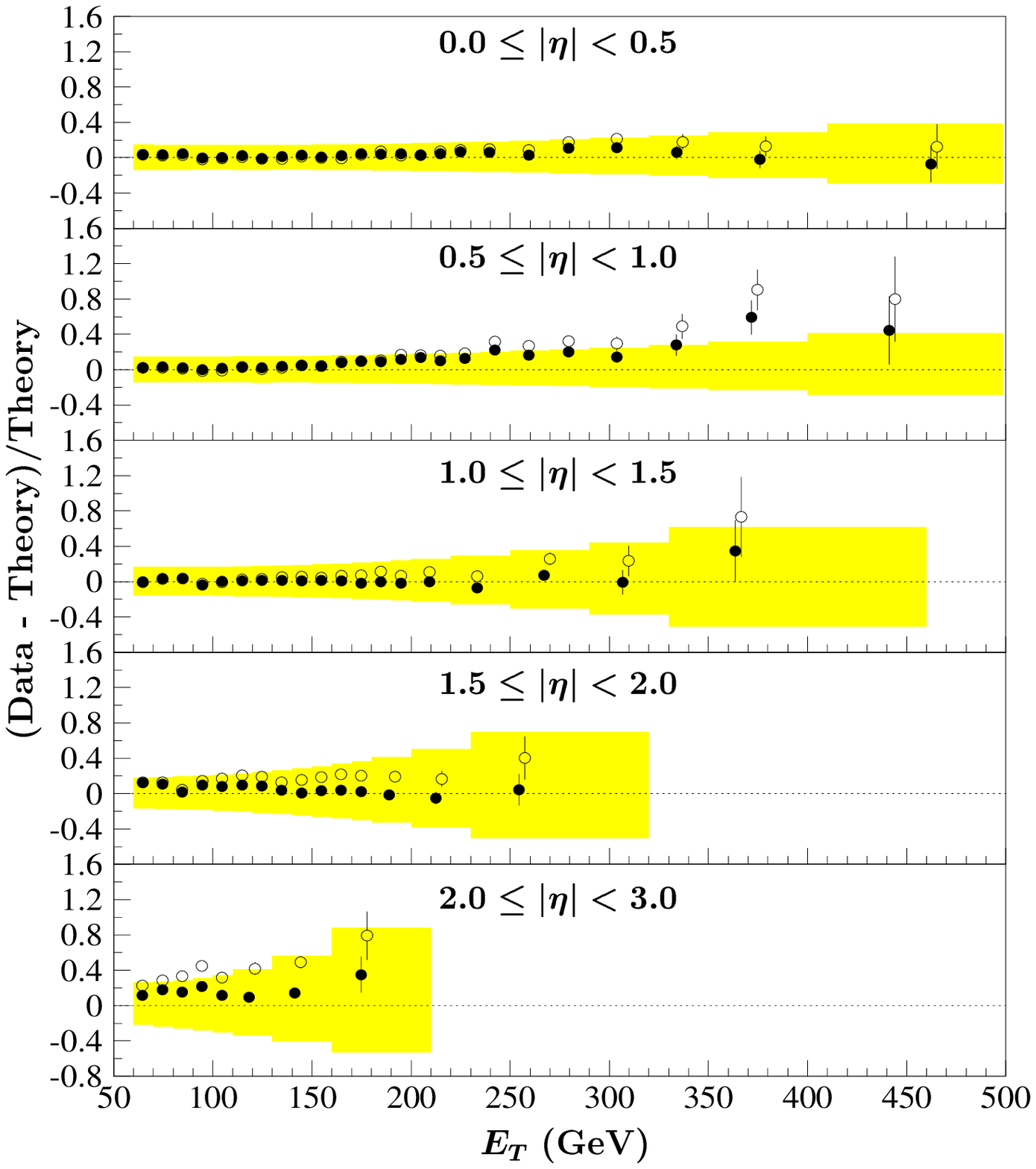}
	    \end{picture}}

  \put(43,-27){\begin{picture}(59,85)
  	      \epsfysize=14.0cm 
  	      \epsfxsize=8.6cm 
	      \epsfbox{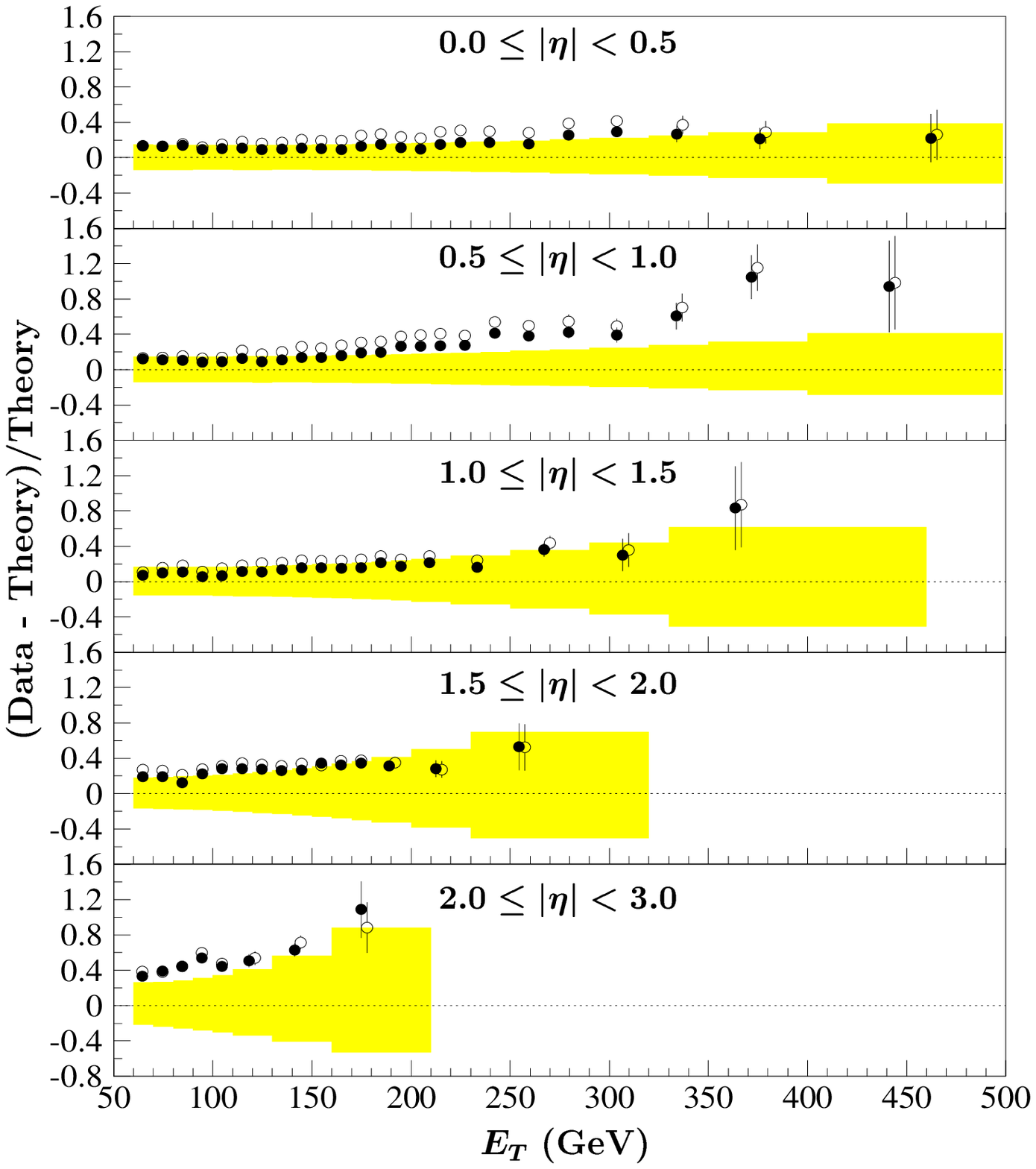}
	    \end{picture}}

  \end{picture}
  \vspace{-0.7cm}
  \caption{Comparisons between the \dzero\ single inclusive jet cross sections 
           and the ${\mathcal{O}}\!\left(\alpha_{s}^{3}\right)$ QCD predictions 
	   calculated by {\sc JETRAD}.
	   Left panel shows comparisons with the CTEQ4HJ ({\boldmath$\bullet$}) 
	   and CTEQ4M ({\boldmath$\circ$}) PDFs while the right panel shows
	   comparisons with the MRSTg$\uparrow$ ({\boldmath$\bullet$}) and MRST 
	   ({\boldmath$\circ$}) PDFs. (The highest \et\ points are offset slightly 
	   for CTEQ4M and MRST.)}
  \vspace{-0.8cm}
  \label{fig:2}
\end{figure}

\vspace{-0.3cm}
\section{Inclusive Jet Cross Section}
\noindent
The final measurements in each of the five \aeta\ regions, along with 
statistical uncertainties, are presented in Fig.~\ref{fig:1}b (tables 
of the measured cross sections can be found in~\cite{myprl}).
The measurement spans about seven orders of magnitude and extends to the 
highest jet energies ever reached.
Figure~\ref{fig:1}b also shows ${\mathcal{O}}\!\left(\alpha_{s}^{3}\right)$ 
theoretical predictions from {\sc JETRAD}~\cite{theory} with renormalization 
and factorization scales set to half of the \et\ of the leading jet and using the 
CTEQ4M PDF.

Left and right panels in Fig.~\ref{fig:2} provide more detailed comparisons 
to predictions on a linear scale for several PDFs (for other PDFs, see~\cite{myprl}).
The error bars are statistical, while the shaded bands indicate one
standard deviation systematic uncertainties.
The theoretical uncertainties due to variations in input parameters 
are comparable to the systematic uncertainties. 
These qualitative comparisons indicate that the predictions are in reasonable
agreement with the data for all \aeta\ intervals.

\renewcommand{\arraystretch}{1.1}
\begin{table}[!ht]
\caption{The $\chi^{2}$, $\chi^{2}$/dof, and the corresponding probabilities
         for 90 degrees of freedom for various PDFs studied.}
\vspace{-0.2cm}
\begin{center}
\begin{tabular}{lccc}
{\bf PDF} & \boldmath $\chi^{2}$ &  \boldmath $\chi^{2}/\mathrm{dof}$ & {\bf Probability} \\ \hline
CTEQ3M            &  121.56  &   1.35  &    0.01   \\ 
CTEQ4M            &   92.46  &   1.03  &    0.41   \\ 
CTEQ4HJ           &   59.38  &   0.66  &    0.99   \\ 
MRST              &  113.78  &   1.26  &    0.05   \\ 
MRSTg$\downarrow$ &  155.52  &   1.73  & $<$0.01   \\
MRSTg$\uparrow$   &   85.09  &   0.95  &    0.63   \\
\end{tabular}
\end{center}
\vspace{-0.5cm}
\label{tab:ratios}
\end{table}
\renewcommand{\arraystretch}{1.3}

To quantify the comparisons, we employ a specially derived and
previously studied $\chi^{2}$ statistic~\cite{myprl,jetsPRD} employing
all $90$ \peta--\et\ bins in this measurement, including correlations in
\et\ as well as in \peta.
For all PDFs we have considered, Table~\ref{tab:ratios} lists the $\chi^{2}$, 
$\chi^{2}$/dof, and the corresponding probabilities for 90 degrees of freedom (dof).
We have verified that the variations of correlation coefficients within the range 
of their uncertainties give a similar ordering of the $\chi^{2}$, hence a similar
relative preference of PDFs.
The absolute values of $\chi^{2}$ and associated probabilities vary somewhat with 
variations in the correlations in \et\ and, to a much lesser extent, with variations 
of correlations in \peta.
The theoretical predictions are in good quantitative agreement with the experimental
results.
The data indicate a preference for the CTEQ4HJ, MRSTg$\uparrow$, and CTEQ4M PDFs.
The CTEQ4HJ PDF has enhanced gluon content at large $x$, favored by previous measurements 
of inclusive jet cross sections at small \peta, 
relative to the CTEQ4M PDF.
The MRSTg$\uparrow$ PDF includes no intrinsic parton transverse momentum and therefore
has effectively increased gluon distributions relative to the MRST PDF.
This measurement provides the world's best constraint on the gluon distribution at high 
$x$ and is being included in the new global PDF fits by the MRST and CTEQ Collaborations.

\end{document}

%% file: definitions.tex
\renewcommand{\arraystretch}{1.3}
\newcommand{\dzero}{\mbox{D\O}}

\newcommand{\z}{\mbox{$z$}}

\newcommand{\et}{\mbox{$E_{T}$}}

\newcommand{\peta}{\mbox{$\eta$}}				
\newcommand{\aeta}{\mbox{$|\eta|$}}				
\newcommand{\ipb}{pb$^{-1}$}

\newcommand{\met}{\mbox{${\hbox{$E$\kern-0.63em\lower-.18ex\hbox{/}}}_{T}$}}
\newcommand{\metvec}{\mbox{${\hbox{$\vec{E}$\kern-0.63em\lower-.18ex\hbox{/}}}_{T}\,$}}
\newcommand{\metx}{\mbox{${\hbox{$E$\kern-0.63em\lower-.18ex\hbox{/}}}_{x}\,$}}
\newcommand{\mety}{\mbox{${\hbox{$E$\kern-0.63em\lower-.18ex\hbox{/}}}_{y}\,$}}

\newcommand{\etaphi}{\mbox{$\eta-\varphi$}}

\newcommand{\etal}{{\it et al.}}
\newcommand{\ppbar}{\mbox{$p\overline{p}$}}

\newcommand{\mycs}{\mbox{$d^{\,2}\sigma/(d\et d\eta)$}}

\def\D0{D\O}
\def\ETmiss{{\rm {\mbox{$E\kern-0.57em\raise0.19ex\hbox{/}_{T}$}}}}

\def\simge
{\mathrel{\rlap{\raise 0.53ex \hbox{$>$}}{\lower 0.53ex \hbox{$\sim$}}}}
\def\simle
{\mathrel{\rlap{\raise 0.53ex \hbox{$<$}}{\lower 0.53ex \hbox{$\sim$}}}}

\def\ETmiss{\mbox{${\hbox{$E$\kern-0.5em\lower-.1ex\hbox{/}\kern+0.15em}}_{\rm T}$}}

\def\1800{$\sqrt{s}=1800$ GeV}
\def\630{$\sqrt{s}=630$ GeV}